\documentclass[sigconf]{acmart}

\usepackage{stfloats} %

\usepackage[font=scriptsize]{subcaption}

\usepackage{url}
\usepackage{caption} %
\usepackage{graphicx} %
\usepackage{booktabs}
\usepackage{multirow}
\usepackage{adjustbox}  %
\usepackage{algorithm}
\usepackage{algorithmic}
\usepackage{amsmath}
\usepackage{natbib}
\usepackage{xspace}

\usepackage{color}
\newcommand{\guohao}[1]{{#1}}	%
\newcommand{\review}[1]{{ #1}}	%
\newcommand{\YS}[1]{{\color{green} #1}}	%

\newcommand{\var}{{\rm Var}}

\newcommand{\f}{{\bf f}}
\newcommand{\g}{{\bf g}}

\newcommand{\s}{{\bf s}}

\newcommand{\w}{{\bf w}}
\newcommand{\x}{{\bf x}}
\newcommand{\y}{{\bf y}}
\newcommand{\z}{{\bf z}}

\renewcommand{\H}{{\bf H}}
\newcommand{\I}{{\bf I}}

\newcommand{\K}{{\bf K}}
\renewcommand{\L}{{\bf L}}

\newcommand{\Dcal}{\mathcal{D}}

\newcommand{\X}{{\bf X}}

\newcommand{\balpha}{\boldsymbol{\alpha}}

\newcommand{\bomega}{\mathbf{\omega}}

\newcommand{\btheta}{\boldsymbol{\theta}}

\newcommand{\1}{{\bf 1}}

\newcommand{\ben}{\begin{enumerate}}
\newcommand{\een}{\end{enumerate}}

\newcommand{\argmin}{\operatornamewithlimits{argmin}}
\newcommand{\argmax}{\operatornamewithlimits{argmax}}

\newcommand{\ie}{{i.e.,}\xspace}
\newcommand{\eg}{{e.g.,}\xspace}

\newcommand{\EE}{\mathbb{E}}

\newcommand{\cmt}[1]{}

\newcommand{\RR}{\mathbb{R}}

\renewcommand{\x}{\textbf{\textit{x}}}  %
\renewcommand{\w}{\textbf{\textit{w}}}  %
\renewcommand{\f}{\textbf{\textit{f}}}  %
\renewcommand{\y}{\textbf{\textit{y}}}  %

\def\Eqref#1{Eq.~\eqref{#1}}

\def\Figref#1{Fig.~\ref{#1}}
\usepackage{xspace}
\newcommand{\ours}{ASDK\xspace}
\AtBeginDocument{%
  \providecommand\BibTeX{{%
    \normalfont B\kern-0.5em{\scshape i\kern-0.25em b}\kern-0.8em\TeX}}}

\setcopyright{none}
\renewcommand\footnotetextcopyrightpermission[1]{} % removes footnote with conference information in first column
\begin{document}
\YS{}
\bibliographystyle{unsrt}
\title{Deep Kernel Learning with Probability Entropy for High Dimensional Yield Estimation}
\title{High Dimensional Yield Estimation using Deep Feature and Convolution Batch Acquisition}
\title{High Dimensional Yield Estimation using Shrinkage Deep Features and Convolution Batch Acquisition}
\title{High-Dimensional Yield Estimation using Shrinkage Deep Features and Maximization of Integral Entropy Reduction}

\author{Shuo Yin}
\authornote{Both authors contributed equally to this research.}
\affiliation{%
  \institution{School of Integrated Circuit Science and Engineering, Beihang University}
  \city{Beijing}
  \country{China}}
\email{18231082@buaa.edu.cn}

\author{Guohao Dai}
\authornotemark[1]
\affiliation{%
  \institution{College of Mechatronics and Control Engineering, Shenzhen University}
  \city{Shenzhen}
  \country{China}}
\email{daiguohao2019@email.szu.edu.cn}

\author{Wei W. Xing}
\authornote{Corresponding author.}
\authornote{Also affiliated with Beihang Hangzhou Innovation Institute Yuhang, Hangzhou, China}
\affiliation{%
  \institution{School of Integrated Circuit Science and Engineering, Beihang University}
  \city{Beijing}
  \country{China}}
\email{wxing@buaa.edu.cn}

\begin{abstract}
Despite the fast advances in high-sigma yield analysis with the help of machine learning techniques in the past decade, one of the main challenges, the curse of ``dimensionality'', which is inevitable when dealing with modern large-scale circuits, remains unsolved. 
To resolve this challenge, we propose an absolute shrinkage deep kernel learning, \ours, which automatically identifies the dominant process variation parameters in a nonlinear-correlated deep kernel and acts as a surrogate model to emulate the expensive SPICE simulation.
To further improve the yield estimation efficiency, we propose a novel maximization of approximated entropy reduction for an efficient model update, which is also enhanced with parallel batch sampling for parallel computing, making it ready for practical deployment.
Experiments on SRAM column circuits demonstrate the superiority of \ours over the state-of-the-art (SOTA) approaches in terms of accuracy and efficiency with up to \guohao{10.3x} speedup over SOTA methods.

\end{abstract}


\maketitle

\section{Introduction}
As semiconductor fabrication technology improves by shrinking down its scale to nano-meter, the negative effect of the process variance, \eg doping fluctuation, intra-die mismatches, and threshold voltage variation, arises and causes yield reduction.
For circuits with cells replicated millions of times (\eg SRAM), extremely small circuit failure probability (usually smaller than $10^{-6}$) must be considered to provide a robust design against fabrication process variations, which forms the yield analysis problem.

Monte Carlo (MC) analysis is generally considered the gold standard for yield analysis in industry and academia.
To provide a reasonably accurate yield estimation, MC requires a large number (usually millions) of SPICE simulations, making it infeasible for modern yield problems, \eg yield estimation for SRAM array with more than 500 independent process variation parameters.
Taking a 32M SRAM with a 97\% yield rate for an example, the yield of the bit cells needs to exceed 99.9999\%\cite{NBL-assisted}. Using MC, more than $1\times10^{6}$ MC samples are required to ensure accuracy.
Therefore, academia and industry rely on other approximated yield estimation algorithms to reduce the overall time cost of the repeated simulations.

Instead of drawing samples randomly as in MC, Importance sampling (IS) based approaches draw samples according to a constructed distribution shifted to the likely-to-fail regions. For example, \cite{HSCS} shifts the sampling mean to the min-norm points of each failure region. \cite{AIS} utilizes an adaptive resampling scheme to keep the sample mean updated. \cite{FMCE} combines variance reduction techniques with importance sampling and gate delay model. The convergence of probability estimation can be accelerated because the failure event is more likely to be drawn around the likely-to-fail regions.  However, These IS-based methods can only find the nearest failure region, whereas the other failure regions are ignored, leading to low efficiency or even low accuracy when the number of samples is not sufficiently large.
Surrogate-modeling-based methods construct a surrogate model to approximate the circuit simulators, based on which the yield is estimated.
For example, \cite{AMSV} evaluates the circuit through a Gaussian process regression model, whereas \cite{LRTAsurrogate} utilizes a polynomial chaos expansion model with low-rank tensor approximation to emulate the system performance function.
To further improve a surrogate model's efficiency, the training data is added sequentially based on the current estimations \cite{LRTAsurrogate} rather than just relying on pre-sampling inputs.

It is also possible to combine the IS- and surrogate-based approaches.
\cite{AMSV} uses an RBF neural network to fit the simulation results with an optimal-mean-shift-vector and proposes a heuristic algorithm to find candidates to conduct subsequent experiments.
However, high-dimensional yield estimation is still a challenge for both industry and academia. 
For the surrogate-based method, the data-driven surrogate model (being Gaussian process or deep learning) does not scale well with the dimensionality of the input space due to the ``curse of dimensionality'' where the number of training data to cover the domain grows exponentially.
Similarly, the IS-based method requires a large number of simulations to obtain the likely-to-fail region, which suffers the same exponential growth with the increase of dimension.

To tackle the high-dimensional yield estimation challenge, 
We propose a novel deep kernel learning surrogate model with non-linear feature selection to capture the black box function between input and output; a parallel sampling scheme is also proposed to update the model efficiently.
The novelty of our work includes:
(1) We propose shrinkage deep features to enable the widely used GP surrogate for high-dimensional process variation inputs. Specifically, we introduce an absolute shrinkage in the reproducing kernel Hilbert Space (RKHS) via the Hilbert-Schmidt independence criterion (HSIC) to select the key features that dominate the input-output mapping, making a GP easier to train, robust against overfitting, and efficient to update with any acquisition functions.
(2) We harness the rich model capacity of a deep kernel learning Gaussian process as our surrogate model to capture the complex black-box function of the process variation parameters and their circuit performance metric under a SPICE simulator. 
(3) We proposed a scalable parallel batch strategy to enable massive parallel model updates, which takes advantage of the high dimensionality and turns the ``curse of dimensionality'' into a ``blessing of dimensionality''.
(4) The empirical study shows that \ours is up to \guohao{10.3x faster} and more accurate than the SOTA methods.

\section{BACKGROUND}
\subsection{Problem Definition: Rare Event Analysis}
Define $\x = [x^{(1)},x^{(2)},\cdots,x^{(d)}]^T \in X$ as the variational parameters, which explains the inevitable random variations of a manufacturing process when conducting a SPICE simulation with given design parameters, \eg transistor widths and lengths, resistance values, capacitance values, and bias voltages and currents.
Without loss of generality, $\x$ is assumed independent Gaussian distributed after normalization, \ie 
\begin{equation}
    p(\x) = \prod_i^d \exp \left(- (x^{(i)})^2 /2 \right)/{\sqrt{2 \pi}}.
\end{equation}
For a particular $\x$ (bare in mind that we cannot control the value of $\x$),
the circuit performance metric $z_k$, \eg amplifier gain and memory read/write time, can be considered as a function $z_k=f_k(\x)$. 
When all $K$ metrics are smaller/larger than some predefined threshold $\z^0$, \eg $z_k \leq z^0_k$ for $k=1,\cdots,K$, the circuit with the corresponding parameters $\x$ is considered a qualified design. 
Otherwise, it is a failure case.
Let's use a compact notation $\z = \f(\x)$ to denotes the process, where $\f$ includes all SPICE simulations and other necessary calculations.
Putting this into a strict formulation, 
for a specific design circuit, the circuit failure probability (equivalently the yield rate) $P_f$ is defined
as 
\begin{equation}
    \label{eq:yield x}
    P_f \triangleq \int_\mathcal{X} I(\f(\x)) p(\x) d \x,
\end{equation}
where $I: \RR^k \rightarrow \{0,1\}$ is the indicator function of whether a performance metric passes the predefined criteria.
The integration is challenging as it does not admit a closed-form solution in general.
In practice, we can generate $N$ samples from $p(\x)$ and approximate $P_f$ by
$  {P}_f \approx \frac{1}{N} \sum_{i=1}^N I( \f (\x_i)),$
which become exact when $N \rightarrow \infty$.
Nevertheless, the computation of ${P}_f$ is highly time-consuming as the number of simulations $N$ required to finish the integral is large particular for large $d$, where each simulation can take hours to finish.

\subsection{Surrogate Model} 
To avoid frequent calls to the expensive SPICE simulators and possibly other calculations, we can use a data-driven surrogate model $\g(\x)$ to approximate $\f(\x)$ and use it to provide a quick estimation for any $\x \in \mathcal{X}$.
There are many choices for the surrogate model for the applications, such as RBF neural networks~\cite{AMSV}, polynomial chaos expansion~\cite{LRTAsurrogate}, and Gaussian process~\cite{AMSV}.
The different methods have their own unique characteristics for specific scenarios.
In general, with a surrogate model, we can approximate the yield using 
$\hat{P}_{f}=\frac{1}{N}\sum_{i = 1}^{N}I(\g(x_{i})),$
where $\x_i$ can be obtained cheaply by sampling from the process variation distribution $p(\x)$. The computation is fast because executing $\g(x_{i})$ is computationally cheap once trained.
Note that $\g(\x)$ is a approximation of $\f(\x)$. Thus, $\hat{P}_{f}$ is an approximation of $P_f$. The accuracy of $\hat{P}_{f}$ depends on the $\g(\x)$, which relies on the collected training dataset $D$.
Certainly, we can use the design of experiment (DoE) to generate training inputs using Latin hypercube sampling (LHS) or Sobol sequence. 
However, this can be of low efficiency, particularly for the yield problem where only a few critical regions matters. 
Instead, we can use a sequential update scheme to update the surrogate model such that the surrogate always gets the best update in terms of reducing its error and uncertainty of $P_f$. This is in line with the theory of Bayesian optimization (BO), where to goal is to approach the global optimal by proposing a sequence of \review{query} points.
Following BO, we call the function measuring the contribution of a candidate point $\x$ an acquisition function (Acq).
The sequential model updating can be formulated,
$\x_* = \argmax_{x \in \mathcal{X}} \mathrm{Acq}(\x|D),$
where $\x_*$ indicates the best candidates for the currently available data $D$, which is then updated using $\x_*$ and its corresponding performance metric. 
Thanks to the acquisition function, we can find the failure region more efficiently without wasting our simulation in unnecessary regions,
resulting in a speedup in estimating the yield.

\section{Research Methods}
In this section, we present \ours based on the deep kernel learning Gaussian process, feature selection, adaptive updating, and parallel acceleration for the yield estimation. {The code is available on Github\footnote{https://github.com/SawyDust1228/HSIC-DKL-Yield-Estimation}.}

\subsection{Deep kernel Learning Gaussian process}
Gaussian process (GP) is a common choice as the surrogate model for design space exploration tasks due to its model accuracy and flexibility with uncertainty quantification. 
By giving the set of the process variation parameters and their circuit performance metric under the SPICE simulation, we aim to capture the black-box function of the process variational parameters and their circuit performance metric.

Suppose that we have a set of $N$ yield observations $\{y_i = f(\x_i) + \epsilon_i\}_{i=1}^{N}$, where the noise $\epsilon$ is caused by the numerical error of a simulator and is assumed normally distributed.
A GP model places a prior distribution over the function $f$ as $ f(\x)|\theta \sim \mathcal{GP}(\mu(\x), k(\x, \x'|\btheta))$, where $\mu$ is the mean function, and the kernel function $k$ is parameterized by ${\btheta}$.
Usually, the mean function can be assumed zero, i.e., $\mu(\x) \equiv 0$, by centering the data. The kernel function can take many forms, the standard RBF kernel are as 
$k({\x}, {\x}'|\pmb{\theta})=\theta_0\exp(-({\x}-{\x}')^T\mbox{diag}(\theta_1,\hdots,\theta_l)({\x}-{\x}'))$.
With this prior and available data $\y=( y_1,\hdots,y_N)^T$, we can derive the model likelihood
\begin{equation} \label{MLE}
    \begin{aligned}
        \bar{L} 
        = -\frac{1}{2} {\y}^T  (\textbf{K} +\sigma^2\I) ^{-1} {\y} -\frac{1}{2}\ln|\textbf{K} + \sigma^2\I|  - \frac{N}{2} \log(2\pi),
    \end{aligned}
\end{equation}
where $\K=[K_{ij}]$ represent the covariance matrix, in which $K_{ij}=k({\x}_i, {\x}_j)$, $i,j=1,\hdots,N$, and $\sigma^2$ denotes the variance of the noise $\epsilon$.
The hyperparameters $\pmb{{\theta}}$ are normally obtained from point estimates by maximum likelihood estimate (MLE) w.r.t. ${\btheta}$.

Recently, deep neural networks have achieved great success in many areas because of their remarkable capacity for feature extraction. \cite{DKL-Paper} combines the non-parametric flexibility of a kernel function with the powerful model capacity of the deep neural networks, which significantly improves the performance of a general GP. 
Despite its success, the underlying ideas are rather simple. It essentially redefines the kernel as
\begin{equation}
    \hat{k}(\x, \x') = k(\phi(\x, \w), \phi(\x', \w)),
\end{equation}
where $\phi(\x, w)$ is a deep neural network (for instance, a multi-layer perception (MLP) with multiple hidden layers) parameterized by weights $\w$ and $k(\cdot,\cdot)$ is any valid kernel function, \eg RBF kernel.

\subsection{Shrinkage Deep Feature Selection}
In general, the process variation parameters quantify the process corner and other factors during the fabrication of a circuit, \eg threshold voltage, channel length modulation effect, and bulk effect.
Usually, the same type of variational variables is applied to each transistor (and/or other crucial elements in the circuit). 
Due to the large number of transistors a practical circuit can have, we will end up with a large number of variational parameters, \eg 1000, causing the ``curse of dimensionality'' issue, which makes both the integration and surrogate fitting extremely challenging.

Fortunately, previous research \cite{jiangwei} reveals that not all variational parameters are equally important. 
In fact, circuit performance is dominated by several critical transistors, whereas the other transistors have little influence, especially in the case where the circuit has a special symmetric design to alleviate the process variation.
This finding makes ``dimension reduction'' possible to reduce the input dimension such that only the key parameters are preserved. 
Note that this is more of a ``feature selection'' rather than dimension reduction because the inputs are {fully independent}, %
indicating that no dimension reduction techniques, \eg PCA and KPCA, can achieve any success.

Let $\{x_i, y_i\}_{i=1}^N$ denotes the collection of all variation parameters and corresponding performance metric $y \in \RR^D$.
The surrogate model aims to approximate $\f(\x)$ with a small $N$ and a large $D$, which can easily lead to overfitting.
As mentioned above, only a few variation parameters matter the most.
We consider the classic feature selection method, least absolute shrinkage selection operator (LASSO)~\cite{lasso}, which chooses the key feature that best explains the input-output relationship.
Essentially, LASSO casts an optimization problem,
\begin{equation}
    \label{eq:lasso}
    \argmin_\alpha \frac{1}{2}||\y-\X\balpha||^{{2}} + \lambda||\balpha||^{1},
\end{equation} 
where $\balpha \in \RR^D$ is a weight vector determining the contribution of each dimension of $\x$, $\lambda$ is the penalty factor; $||\cdot||^2$ denotes the L2 norm and $||\cdot||^1$ the L1 norm. 
This formulation is effective {because its gradient} w.r.t $\alpha^{(i)}$ (which is the i-th element of $\balpha$) is a constant, which will push the value of $\alpha^{(i)}$ towards zero unless it contributes significantly to reducing the data fitting loss of the first term.

Despite its elegance and effectiveness, we can immediately tell the limitation of LASSO is that it is based on a linear model, whereas the yield analysis problem we are facing is usually nonlinear.
An ordinary LASSO will ignore the nonlinear connections and only focus on the linear ones, leading to a poor feature selection. 
Since a GP essentially relies on the reproducing kernel Hilbert space (RKHS) to build the connection between the input and output, we can naturally extend the original LASSO for the RKHS, which leads us to the Hilbert-Schmidt independence criterion (HSIC) proposed by \citep{HSIC-Lasso}. 
Specifically, we require the output correlation measured by the kernel function $k(\cdot,\cdot)$ can be captured via the sum (with weight factor $\alpha$) of the correlation matrix of each input dimension.
The weight factor $\alpha$ is then contrasted using the LASSO trick. 
For a more rigorous derivation, the readers are referred to \citep{HSIC-Lasso}.
In practice, we also apply double centering for the kernel matrix to ensure stability and solve the optimization problem:
\begin{equation}
    \label{eq:HSIC lasso}
    \argmin_\alpha \frac{1}{2}||\widetilde{\L} - \sum_{d = 1}^{D} \K^{(d)} \alpha^{(d)} ||^{2} + \lambda||\balpha||^1
\end{equation}
where $\widetilde{\L} = \H \K_y \H^T$, with $\H=\I-\frac{1}{N}\1 \1^T$ being the centering matrix;
$[\K_y]_{ij}=k(\y_i, \y_j)$ and $[\K^{(d)}]_{ij}=k(\x^{(d)}_i, \x^{(d)}_j)$ are kernel matrix given the target values and d-dimension of $\x$; $k(\cdot, \cdot)$ is the kernel function with default hyper-parameters.
The key result of this algorithm is the weight vector $\balpha$, which indicates the importance of each input dimension that dominates the input-output mapping in the RKHS, which is encoded in our GP.

\subsection{Maximum Integral Entropy Reduction}
In Bayesian Optimization, the acquisition usually uses expected improvement (EI), predictive improvement (PI), and upper confidence bound (UCB) to propose a candidate for the next iteration. However, they are designed for optimization and do not generalize to yield estimation directly.

We notice that the circuit yield is usually very high (equivalently, the failure rate is very low), which means that the passing threshold $\z^0$ is an extreme value compared to most simulation performances, and only a few samples will eventually fail the indication function.
This hints that we can reduce the computing cost of integral by avoiding observation in the region that the simulation performance $z$ will "absolutely" pass the criterion and try to locate the boundary of the failure region. This search scheme is in line with the boundary search method.

Considering a GP $\f(\x)=[f_0(\x), f_1(\x) \hdots f_K(\x)]^T$, the posterior of each $f_i(\x) \sim \mathcal{GP}(m_i(\x), k_i(\x, \x'))$ is also a gaussian distribution with mean $\mu_i(\x)$ and variance $v_i(\x)$. 
The predictive posterior based on the pass threshold $\z^0$ is a Bernoulli distribution with likelihood $l(\x) \triangleq p(\widetilde{I}(\x)=1)$, where $\widetilde{I}(\x) = I(\f(\x))$ indicates the predictive performance,
\begin{equation}
    \label{eq:yield prob}
    l(\x) =\prod_{k=1}^K p\left( \widetilde{f_k}(\x) \geq z_k \right)=
    \prod_{k=1}^K \Phi(\frac{\mu_k(\x) - z_k^0}{v_k(\x)}),
\end{equation}
where $\Phi(\cdot)$ is the cumulative density function (CDF) of a normal distribution.
According to the Poisson binomial distribution, which is the sum of independent yes/no experiments. We derive the approximated integral $\tilde{g}$ in a similar manner,
\begin{equation}
    \label{eq:yield post}
    \begin{aligned}
        \EE \left[\hat{P}_f\right]
        &= \int_\mathcal{X} l(\x)p(\x) d \x, \\
        \var \left[\hat{P}_f\right]
        &= \int_\mathcal{X} l(\x) \left(1-l(\x)\right) p(\x) d \x.
    \end{aligned}
\end{equation}
Because $l(\x)$ is tractable, and $p(\x)$ is simply a diagonal Gaussian, we can compute the integral efficiently using numerical approximations or quasi-MC.
Achieving an accurate yield estimation $\hat{P}_f$ is equivalent to reducing its variance $\var \left[\hat{P}_f\right]$, which, however, ignores the high-order moment and can lead to inferior results.
Instead of reducing the variance, we introduce a probability information entropy for the yield posterior of $\tilde{I}(\x)$, which is the entropy of a Bernoulli distribution,
\begin{equation}
    H(\x) = - l(\x) \log\left( l(\x) \right) - (1- l(\x)) \log\left(1-l(\x)\right).
\end{equation}
We then define the total integral entropy as
\begin{equation}
    \label{eq:IH int}
    IH = \int_{\mathcal{X}} H(\x) p(\x) d\x,
\end{equation}
which indicates the uncertainty of $g(\x)$ based on the surrogate model with current observations $D$.
To reduce the uncertainty of $P_f$, we can propose a candidate base on maximizing the expected integral entropy reduction,
\begin{equation}
  \label{eq:IH opti}
  \begin{aligned}
    \x^* &= \argmax_{\x \in \mathcal{X}} \left( IH(\Dcal) - IH(\Dcal \cup \x) \right) \\
    & = \argmin_{\x \in \mathcal{X}}  IH(\Dcal \cup \x).
  \end{aligned}
\end{equation}
Thus, to get the optimal $\x^*$, we first draw multiple samples from the predictive posterior $f_k(\x|\Dcal)$ for a possible observation performance metric $\z^{(k)}$.
We then combine those samples with our data collection $\Dcal$ and update the posterior as $f_k(\x|\Dcal \cup \{\x,\z^{(k)}\})$, based on which, we can compute the integral entropy $IH(\Dcal \cup \x)$.

\subsection{Parallel Batch Acquisition}
\label{sec:parallel}
In practical yield applications, it is important to allow a parallel acquisition of multiple candidates to unleash the power of the modern cluster center.
In a recent work \cite{LRTAsurrogate}, the authors use a mixed Gaussian distribution to implement a parallel updating scheme. More specifically, they generate a discrete Gaussian sampling distribution around each observed data and use it to sample new candidates. 
However, these new candidates may fall in the same region with a high probability in the input domain, leading to an inefficient sampling strategy.
In this section, we generalize \ours for parallel computing.
More specifically, we would like to propose multiple candidates at each iteration. Formally, we aim to solve the optimization,
\begin{equation}
    \label{eq:IH opti2}
    \begin{aligned}       
        \X^* = \argmin_{\X \in \mathcal{X}} \ IH(\Dcal \cup \x_1 \cup \cdots, \cup \x_{Q}),
    \end{aligned}
\end{equation}
where $\X^*$ indicates the collection of $Q$ ideal query points.
As discussed previously, directly solving this equation is challenging and computationally expensive.
To approximately solve this optimization in a batch fashion, we take advantage of the ``curse of dimensionality'' and turn it into a blessing. 
More specifically, we discover that  
if we start the \ours at multiple far-away initial locations, the final {query} points do not converge to the same locations due to the complex geometry in the high-dimensional space.
Inspired by the Q-batch initialization \cite{q_batch}, we convert finding multiple {query} points into finding multiple far-way initial points for optimization.

Assume that we aim to generate $Q$ \review{query} points.
We first pre-sample $T$ points in the domain, where we ensure $T >> Q$.  We then compute these $T$ points' acquisition score $\s$,
$ \{{s_i} = IH(\Dcal \cup \x_i)\}_{i=1}^T.$
Let $s_{m}$ be the max value in these $T$ scores. Define \review{$\gamma$} as a fraction of the maximum observed value under which we will ignore. 
This coefficient helps us filter out the scores that are lower than $\gamma * s_{m}$.
In cases where the number of satisfying candidates 
is smaller than the \review{query} number $Q$, we use the following equation to relax the threshold $\gamma = (1 - \beta) \gamma_0,$
where $\beta$ is another hyperparameter to relax the filter and \review{$\gamma_0$} is the ratio that does not satisfy the need.
We can repeat this process until we get more than $O$ points ($O>Q$) over the threshold. This idea is similar to the two-stage estimation in \cite{two-stage}.
After we get $O$ samples $\{s_{i}\}_{i=1}^O$ that are larger than the threshold, we use the following equation to get the sample weights:
\begin{equation}
    \omega_{i} = exp(\eta_1 \frac{s_i}{s_{m}} ),
\end{equation}
where $\eta_1$ is a scaling factor. 
Based on the weight $\bomega$, we sample $Q$ initial points sequentially from the $K$ points set.

We then conduct maximum entropy reduction in \Eqref{eq:IH opti} in parallel with the $Q$ initial points with gradient descent, \eg SGD or Adam. A summary of our parallel batch \review{query} scheme is presented in Algorithm \ref{algo2}.

\begin{algorithm}[h]
\caption{\ours Parallel Batch \review{query} Algorithm}
\begin{algorithmic}[1]  \label{algo2}
\renewcommand{\algorithmicrequire}{\textbf{Input:}}
\REQUIRE{number of initial set $T$, number of candidate $Q$, number of filtered samples $O$, fraction coefficient \review{$\gamma$}, relax coefficient $\beta$, weight sample coefficient $\eta_1$.}\\
\STATE{Randomly generate $T$ points in the domain using the Sobel sequence and compute the scores for these points.}
\STATE{Assign the threshold $\alpha s_{max}$ and get $O$ candidate points.}
\STATE{Sequentially sample Q points with weight $ \omega_{i} = \exp(\eta_1 \frac{s_i}{s_{m}})  $}
\STATE{Conduct maximum entropy reduction in \Eqref{eq:IH opti} in parallel with the $Q$ initial points to get $Q$ optimal solutions $\X^*$}
\RETURN{Best candidates $\X^*$}
\end{algorithmic} 
\end{algorithm}

\section{Experiment Results}
In this section, we assess \ours with the SOTA yield estimation algorithms on commonly used high-dimensional benchmark circuits.
We compare \ours to the SOTA yield estimation algorithms, including: (1) LRTA \cite{LRTAsurrogate}, a high-dimensional yield estimation algorithm using the low-rank tensor approximate polynomial chaos expansion model as the surrogate model with a KDE-based adaptive sampling strategy to update the model, (2) HDBO \cite{HDBO}, which uses a random embedding feature reduction method and a Bayesian optimization to find the failure event, (3) HSCS \cite{HSCS}, which applies a clustering algorithm to identify multiple failure regions and uses min-norm-points to resample the failure region. 
\guohao{Bayesian-based approaches like HDBO and \ours are implemented with parallel computing.}
All experiments are performed on a Linux system with AMD 5950x, GTX 3080, and 32GB RAM.

\review{reviewer: Comparison with Bayesian Optimization-based methods is lacking.}

To determine when to stop the yield estimation process, we follow the widely used Figure of Merit (FOM) $\rho$ in the yield estimation literature \cite{HSCS,AIS,LRTAsurrogate} as the stopping criteria. FOM is defined by {$\rho = \sigma_{P_{f}}/{P_{f}},$}
where $P_{f}$ denotes the mean failure probability estimation and $\sigma_{P_{f}}$ the standard deviation of $P_{f}$. 
Following the literature, we set the threshold $\rho_0=0.1$, \ie stopping the yield estimation process when $\rho$ < 0.1. This is equivalent to the stopping criteria of convergence with 90\% confidence interval\cite{HSCS}.

Without loss of generality, we test one circuit metric for each experiment, \ie $\z^0 \in \RR$. We can implement high dimensional circuit metric by changing our GP model to a multi-task GP \cite{multi-task} or just fitting each metric independently.
For the experimental purpose, we set the threshold to fix the yield failure rate to be approximately $10^{-4}$ to reduce overall computation and to emphasize the search of multiple failure regions in the process variational space.
The hyperparameters we mention in this section \ref{sec:parallel} is set as following, $\alpha = 0.3$,  $\beta = 0.9$, $T = 100 \times N$, $\eta_1 = 0.5$.
As for the deep kernel network, we use a three-layer MLP when the dimensionality is smaller than 128 and a four-layer MLP for other cases. The proposed method can be applied to general scenarios by changing the structure of the neural networks.

\subsection{18 Dimensional 6T SRAM Bit Cell}
\begin{figure}
  \centering
  \includegraphics[width=0.27\linewidth]{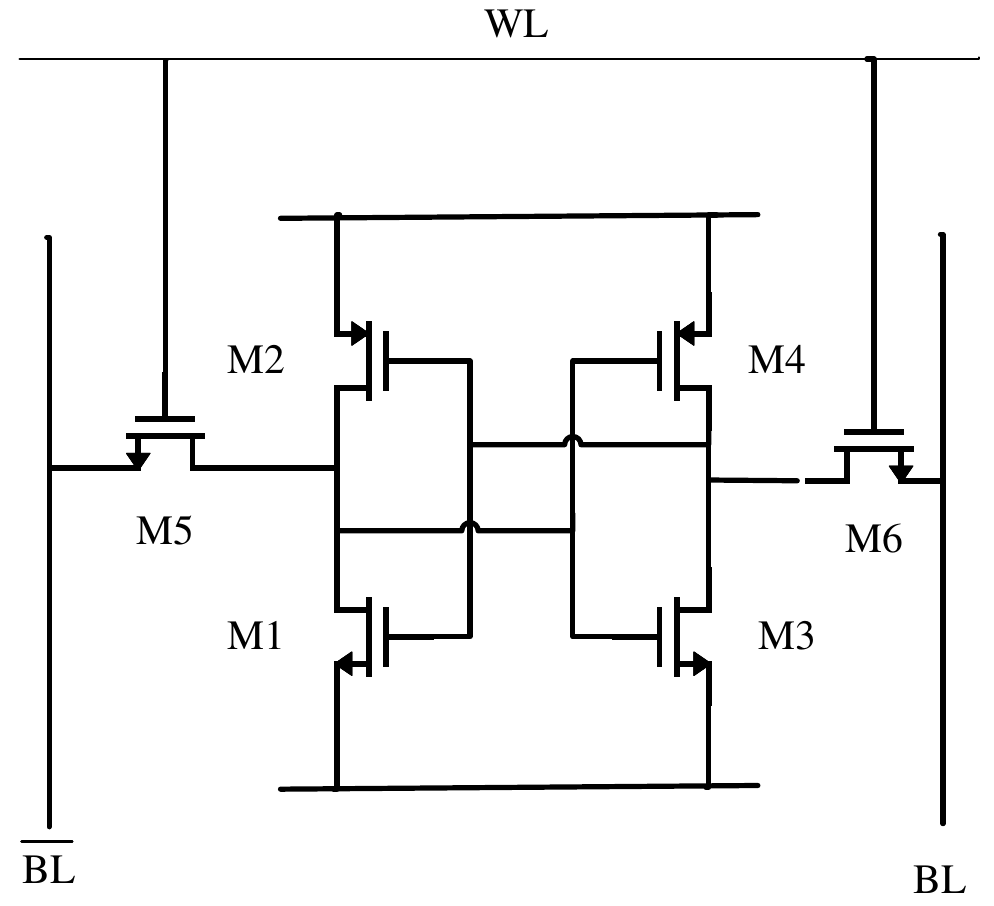}
  \includegraphics[width=0.6\linewidth]{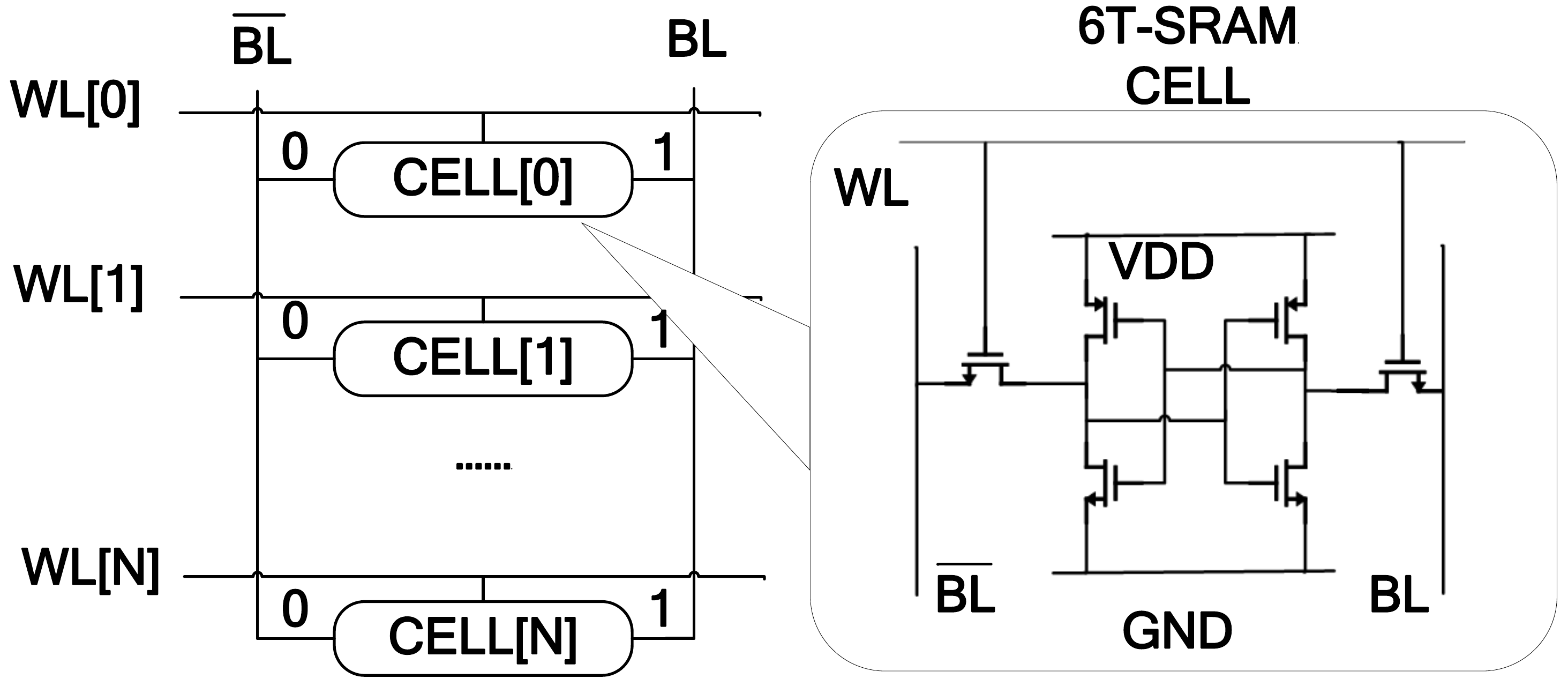}
  \caption{The schematic of 6T-SRAM cell (left) and SRAM array column (right)}
  \Description{}
  \label{6t_SRAM}
\end{figure}

The 6T SRAM bit cell is a basic storage unit of SRAM circuits, which consists of six transistors shown in \Figref{6t_SRAM}.
In the circuit design, M1, M3, M5, and M6 are NMOS transistors, whereas M2 and M4 are PMOS transistors, BL is the bit line, and WL is the word line.
The state of each bit in the SRAM is stored in two cross-connected inverters composed of M1, M2, M3, and M4.
M5 and M6 NMOS are control switches used to control data transmission from the storage unit to the in-place line. 

\begin{figure}[tbp]
	\includegraphics[width=1\linewidth]{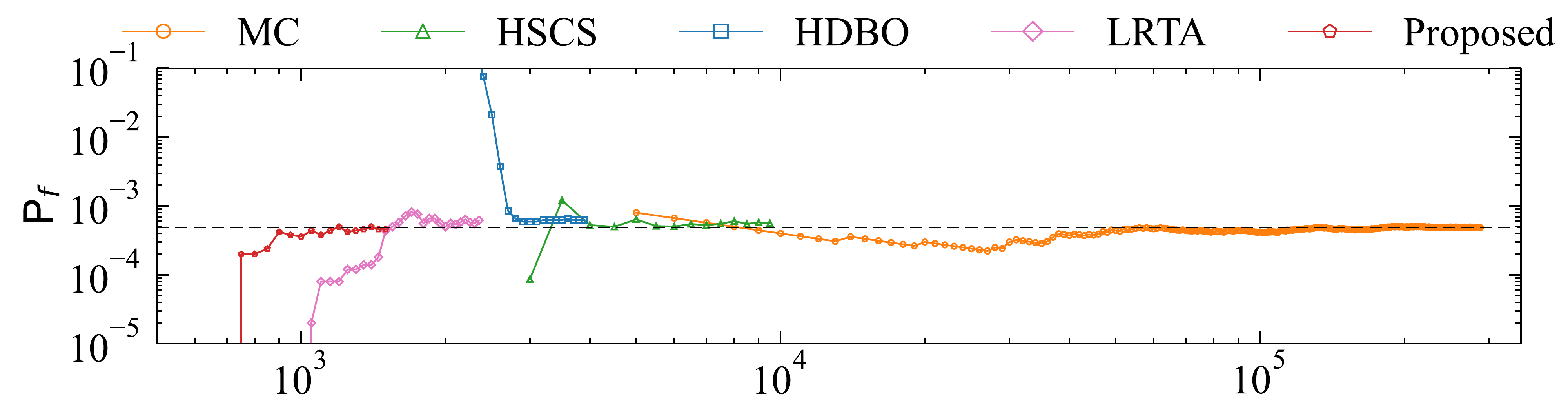}
	\includegraphics[width=1\linewidth]{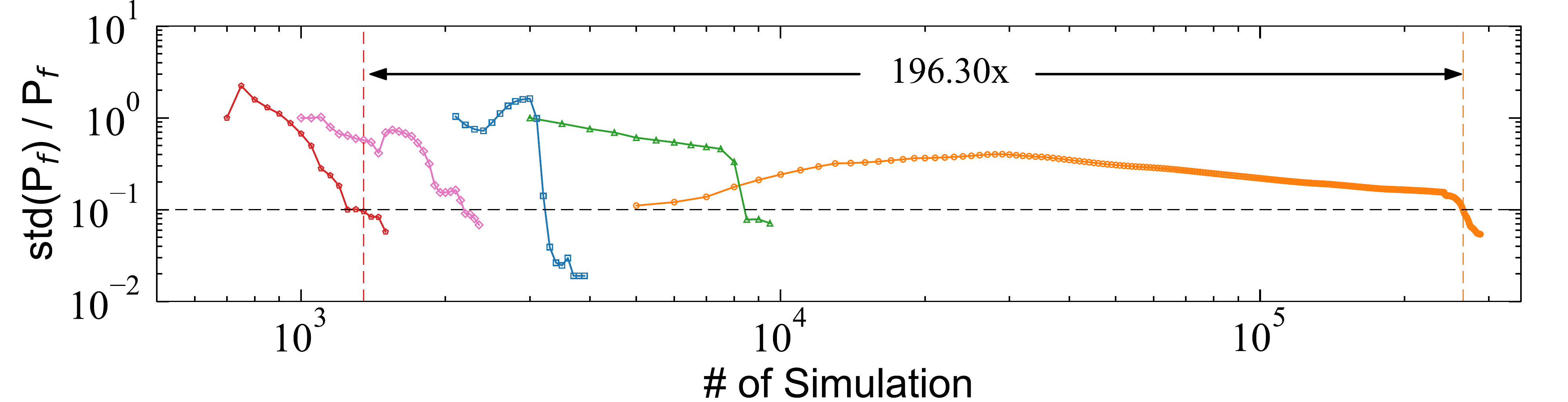}
    \Description{Prob. failure and FOM of 18 dimensional circuit}
    \caption{  $P_f$  and FOM on 18-dimensional 6T SRAM}
  \label{case2_CMR}
\end{figure}

We consider the delay of read/write as the circuit output metric for this experiment, which is a commonly used setup for yield estimation research \cite{LRTAsurrogate,AIS,NON-GAUSSIAN}.
In the setting of this experiment, our deep kernel learning uses a three layers MLP with 200, 100, and 10 hidden neurons and ReLu as the activation function between each layer. 
We use a Matern plus \guohao{a} linear kernel for the GP to keep flexibility and efficiency.
\guohao{Fig. \ref{case2_CMR}} shows the detailed estimation process whereas \guohao{Table \ref{case2Table_CMR}} concludes the experimental results.
It can be seen clearly from the table that \ours outperforms the competitors with a large margin in terms of estimation accuracy and the number of simulations required to achieve convergence, rendering a {196.30x} speedup compared to standard MC and an up to \guohao{6x} speedup to the competitors.
The proposed method is the most accurate with only {4.14\%} relative error w.r.t. the golden truth MC value among the baseline methods.
The MC method reaches the stopping criteria after \guohao{265000} simulations. HSCS, HDBO, and LRTA converge with \guohao{8500, 3500, and 2200} simulations, respectively. 
\guohao{The failure rate of the IS-based HSCS remains small for a long period in the beginning, and a sharp increase in the failure rate will emerge when HSCS discovers all correct failure regions}.
The poor efficiency of {HSCS} is not a surprise, because {an IS-based approach} needs a massive number of pre-sampling data to cover the failure regions. HDBO, LRTA, and the proposed \ours exhibit relatively fast convergence with a 75.71x, a \guohao{120.45x and a 196.30x} speedup because they use surrogates. 
Nevertheless, \guohao{\ours converge to the most accurate fail rate using merely 1350 simulations}. 
{The evolution trend of the yield for surrogate methods is determined by the initial sampling technique and the regression model. 
Due to the different initial fitting techniques to the variation parameter space, the initial failure rates of different surrogate approaches vary.
With the surrogates being updated, the fail rates of LRAT and \ours grow larger gradually, whereas that of HDBO grows smaller. Nevertheless, they all converge to the ground truth with enough simulation runs. }

\guohao{The time consumed in model training (the time spent in simulation run is not included) during the whole yield estimation process of each approach is shown in Table \ref{case2Table_CMR}. MC needs no training. IS-based HSCS spends only 5.61s on training. Surrogate-based methods like HDBO, LRTA, and \ours require more time in model training. Among them, HDBO and \ours obviously need more time to train because their Bayesian optimization is time-consuming. The proposed \ours require the most time to train in this case. However, the train time of the model can be ignored when compared with the time of the simulations.}

\begin{table}
  \caption{  Final $P_f$ estimation on 18-dimensional 6T SRAM}
  \label{case2Table_CMR}
  \begin{tabular}{cccccc}
    \toprule
     & MC & \guohao{HSCS} & HDBO & \guohao{LRTA} & Proposed\\
    \midrule
    Failure prob. & 4.83e-4 & 5.50e-4 & 6.25e-4 & 6.40e-4 & 4.60e-4 \\
    Relative error & Golden & 13.87\% & 29.40\% & 19.46\% & 4.14\% \\
    \# of Sim. & 265000 & 8500 & 3500 & 2200 & 1350 \\
    \guohao{Sim. speedup} & 1x & 31.18x & 75.71x & 120.45x & 196.30x\\
    \guohao{Training time} & N/A & 5.61s & 401.62s & 53.50s & 1537.73s\\
  \bottomrule
\end{tabular}
\end{table}

\subsection{569 Dimensional SRAM Column}
\label{exp3}
{\Figref{6t_SRAM} shows the simplified circuit diagram of the 6T-SRAM bit cell array. Similarly, we use the delay of read/write as the metric. We increase the dimensionality with more cells, which leads to 569 process variation parameters. When the dimension get large, we need to increase our deep network's capacity to deal with the more complex data structure. To this end, 
we use an MLP with {1000, 500, 200, and 20 hidden units} with the hyperparameters remaining unchanged. The number of selected features is set at 120.

The same experimental results are shown in \guohao{ \Figref{case4_CMR} and Table \ref{case4Table_CMR}}. \guohao{The fail rate of HSCS remains very low (6.76e-27) until it finds failure regions near the origin of the coordinates of the variational parameter space. Therefore, the first half of its $P_f$ curve is unable to be plotted in the figures. }
{Surrogate-based methods like HDBO, LRTA, and \ours converge to the ground truth gradually.}
We can see that \ours again achieves the lowest relative error among all methods with the minimum number of simulation runs, rendering its superiority over the SOTA methods. Particularly, compared with the second best method, LRTA, \ours is \guohao{2.9x} more accurate and \guohao{1.3x} faster. The improvement over HSCS is about \guohao{10.3x} faster with a \guohao{3.9x} improvement in accuracy.
As for the training time, LRTA requires 12403s, which makes it the lowest efficient method in this case due to the exponential growth of complexity w.r.t. dimensionality.
Notice that \ours also requires quite a significant training time. However, we believe that this computational cost is weightless than the simulation runs.

\begin{table}
    \caption{
      Final $P_f$ on 569-dimensional SRAM column}
    \label{case4Table_CMR}
    \begin{tabular}{cccccc}
      \toprule
       & MC & \guohao{HSCS} & HDBO & \guohao{LRTA} & Proposed\\
      \midrule
      Failure prob. & 4.70e-4 & 5.92e-4 & 3.87e-4 & 5.60e-4 & 4.39e-4 \\
      Relative error & Golden & 25.96\% & 17.66\% & 19.14\% & 6.60\%\\
      \# of Sim & 928500 & 41500 & 6100 & 5400 & 4000  \\
      \guohao{Sim. speedup} & 1x & 22.37x & 152.21x & 171.94x & 232.13x\\
    \guohao{Training time} & N/A & 108.71s & 1001.73s & 12403.s & 5546.56s \\
    \bottomrule
  \end{tabular}
  \end{table}
  
\begin{figure}[tbp]
	\includegraphics[width=1\linewidth]{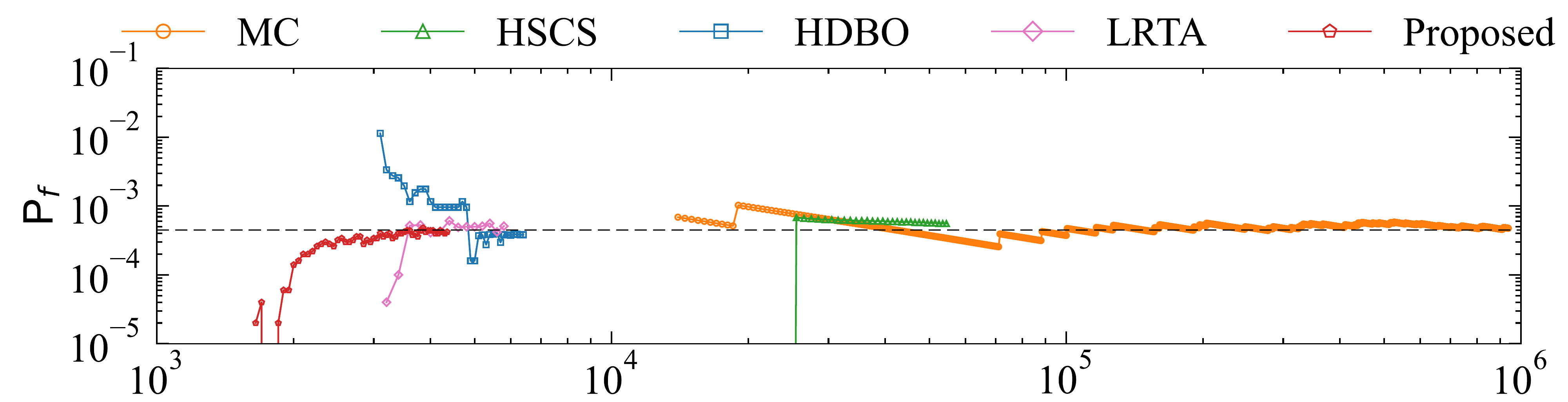}
	\includegraphics[width=1\linewidth]{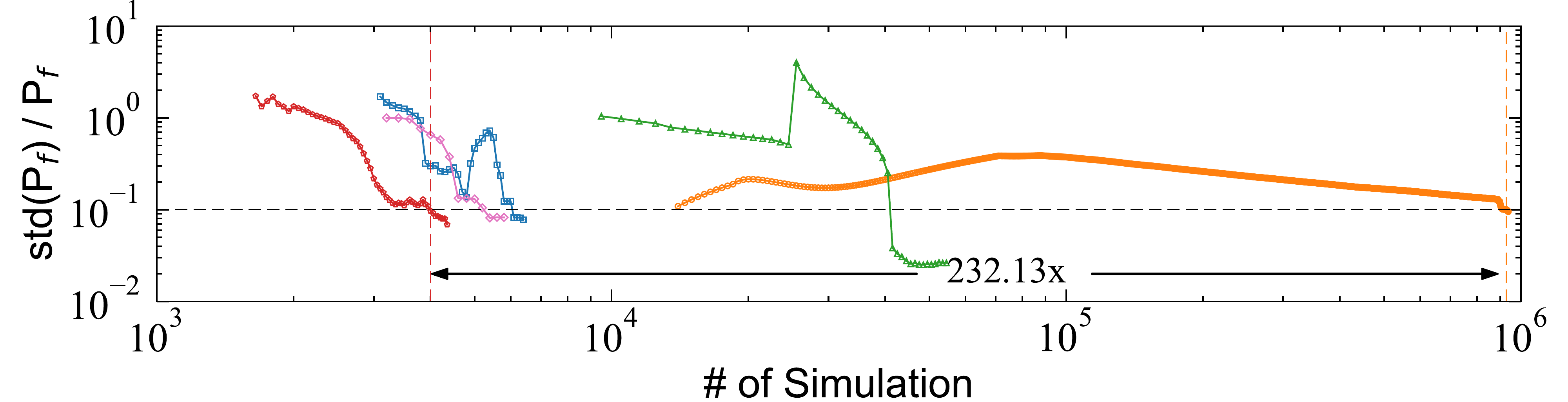}
	\caption{  $P_f$ and FOM on 569-dimensional SRAM column}
  \Description{Prob. failure of 569 dimensional case}
  \label{case4_CMR}
\end{figure}

\subsection{Ablation Study}
\noindent\textbf{Parallel Batch Update Convergence Validation.}
To assess the proposed parallel batch update method,
we conducted the parallel experiment on both previous experiments with different batch sizes, \ie \{20,50,100\} and show the yield in \Figref{parallel2}.
We can see that a parallel run with 100 simultaneous candidates is almost as good as with 20, indicating the scalability of the proposed batch method.

\begin{figure}[tbp]
    \centering
    \includegraphics[width=0.95\linewidth]{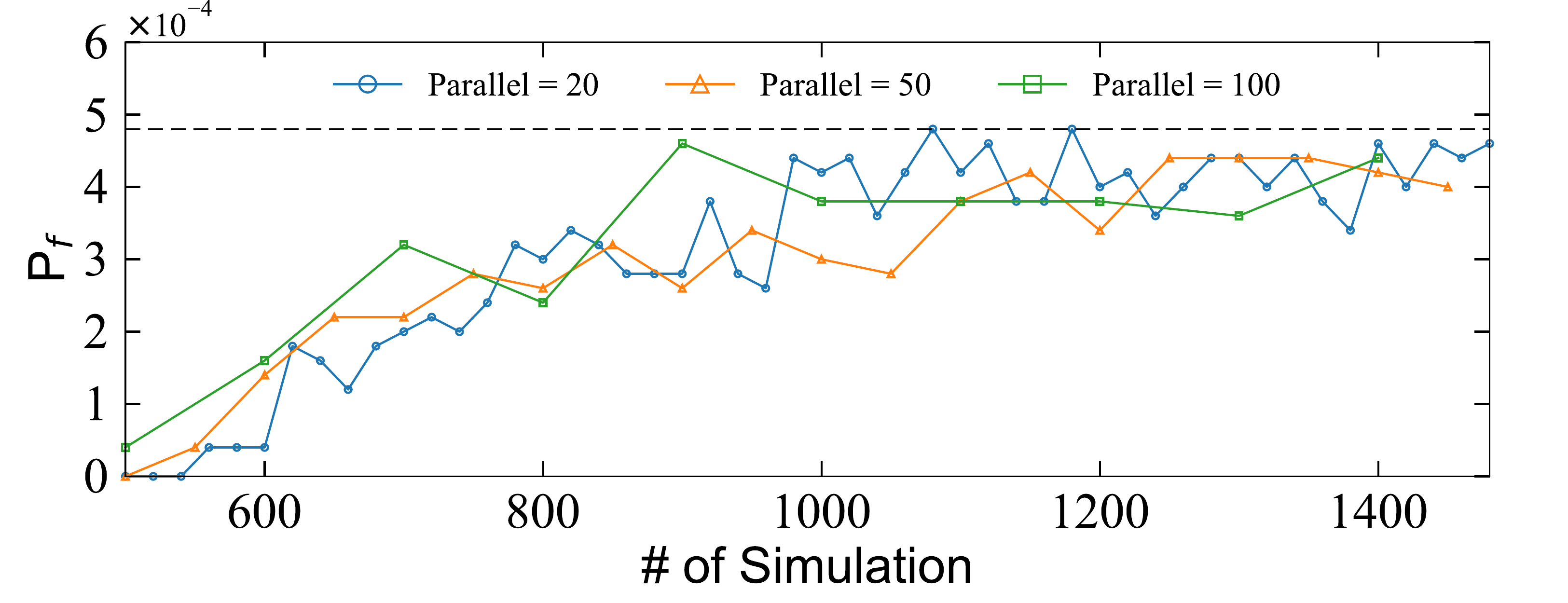}
    \includegraphics[width=0.95\linewidth]{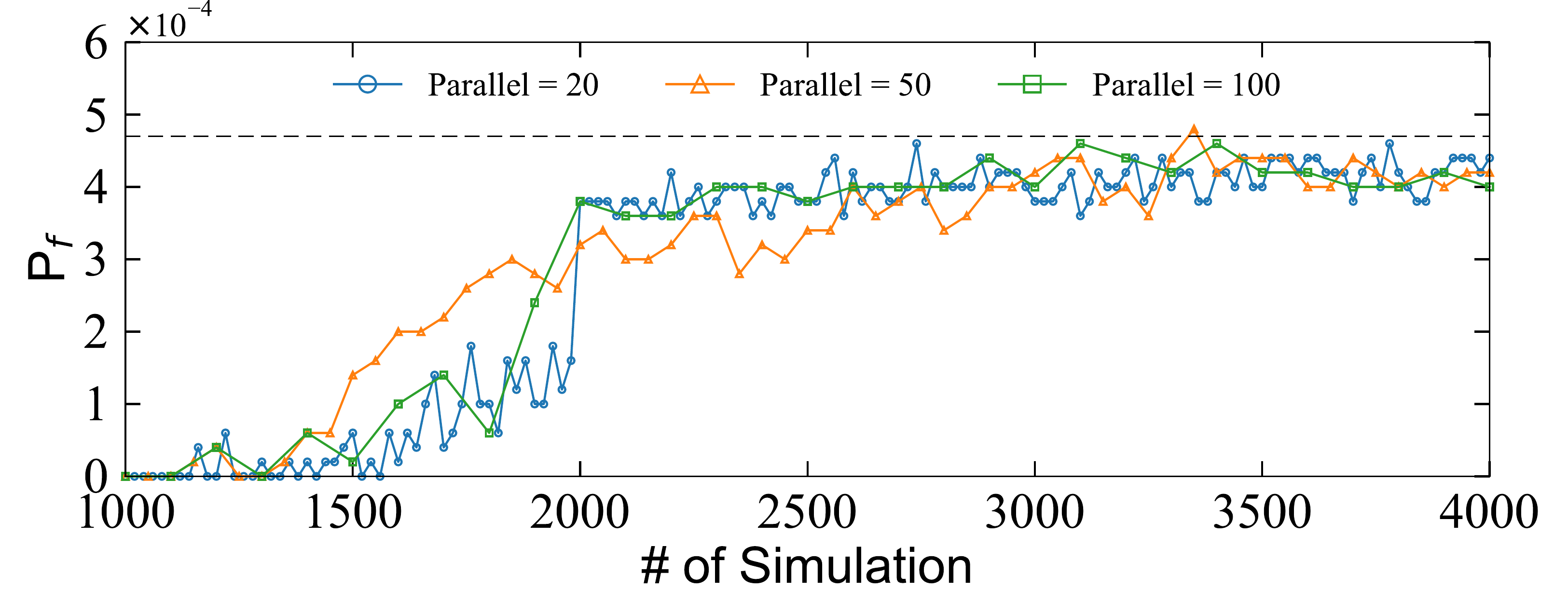}
    \caption{$P_f$ estimation with different batch size}
    \Description{Parrallel experiment of both cases}
    \label{parallel2}
\end{figure}

\noindent\textbf{Maximum Integral Entropy Infill Validation.}
{To demonstrate the superiority of the proposed surrogate updating method of \ours over the commonly used EI, PI, and UCB acquisition functions,
we run \ours with different methods and compare the estimated $P_f$ in Fig. \ref{acquisition} for both previous experiments.
Despite that all methods converge to the ground truth, 
the proposed Maximum Integral Entropy Infill outperforms the competitors in terms of the convergence rate w.r.t the number of simulations in both cases, which is essential in yield analysis to save the expensive simulation cost.
\begin{figure}[tbp]
    \centering
    \includegraphics[width=0.95\linewidth]{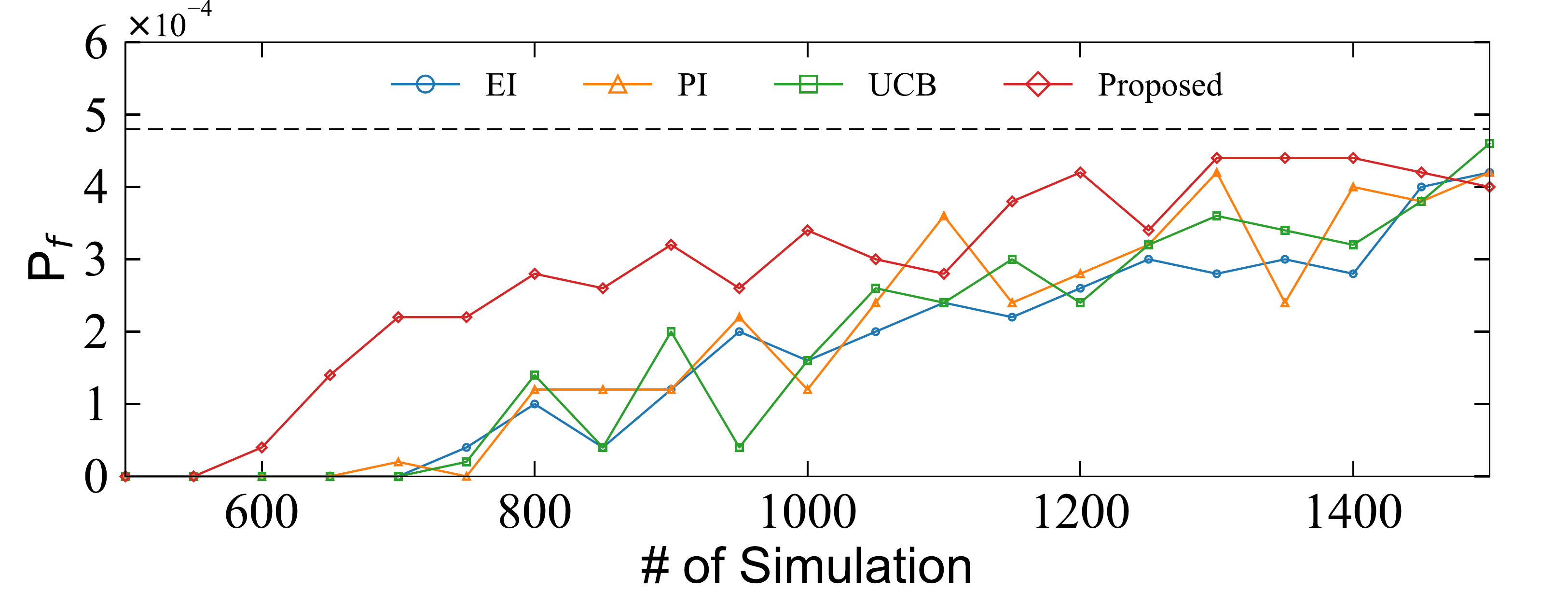}
    \includegraphics[width=0.95\linewidth]{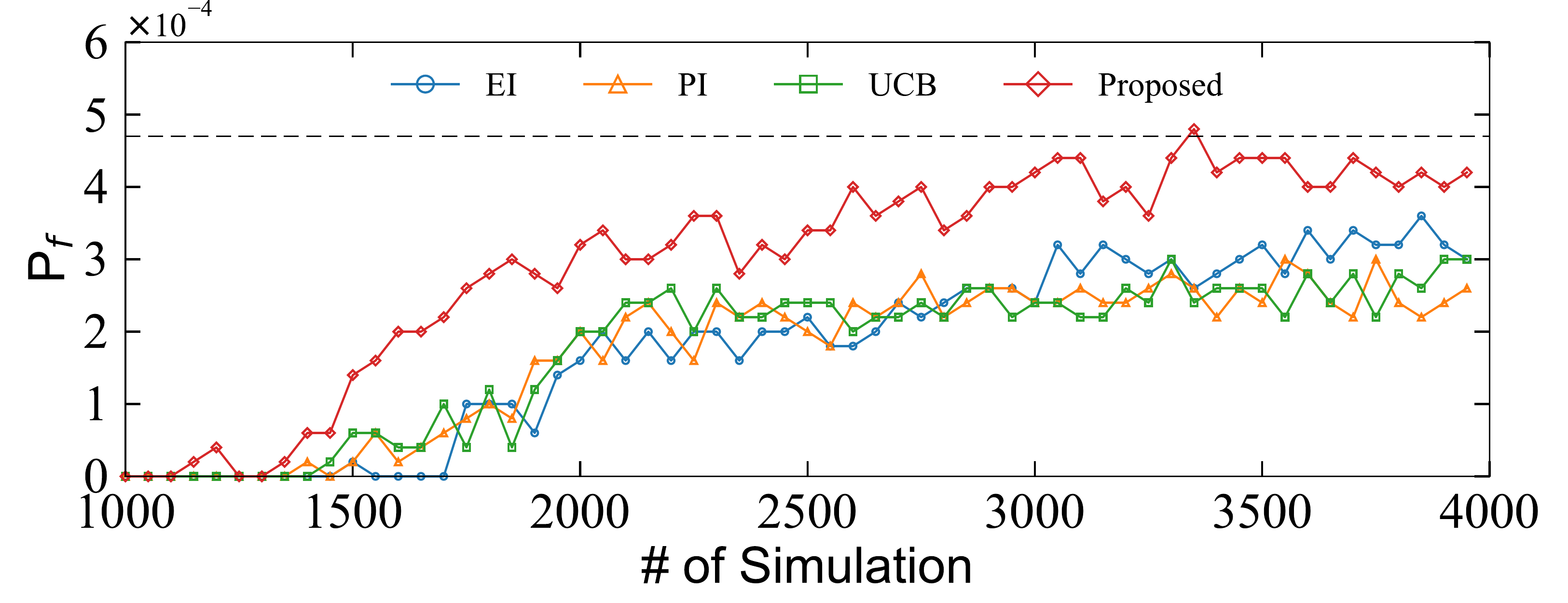}
    \caption{Acquisition function experiment}
    \Description{Acquisition function experiment}
    \label{acquisition}
\end{figure} 

\noindent\textbf{Feature selection Validation.}
{To demonstrate that the feature selection by the HSIC-Lasso algorithm is reliable and efficient, we compare \ours with some traditional feature selection and dimension reduction algorithms, including {Factor Analysis (FA), Principal Component Analysis (PCA), Mutual Information (MI) \cite{jiangwei}, and Random Embedding (RE) \cite{HDBO}.
We first randomly generate 1000 training data through Latin hypercube sampling (LHS).
We run the feature selection/dimension reduction algorithms to reduce the dimensionality to a specified number and assess the model accuracy using RMSE using left-out testing points. 
The experiments are repeated five times with randomly shuffled training and testing data, and the average RMSE against the number of preserved dimensionality is shown in Fig. \ref{dimension}.
We can see very clearly that the implemented HSIC-Lasso is significantly more efficient in preserving the dominant information, leading to a much lower RMSE until the dimension reaches 500. This also partly explains the previous superior performance of \ours.
\begin{figure}[tbp]
    \centering
    \includegraphics[width=0.95\linewidth]{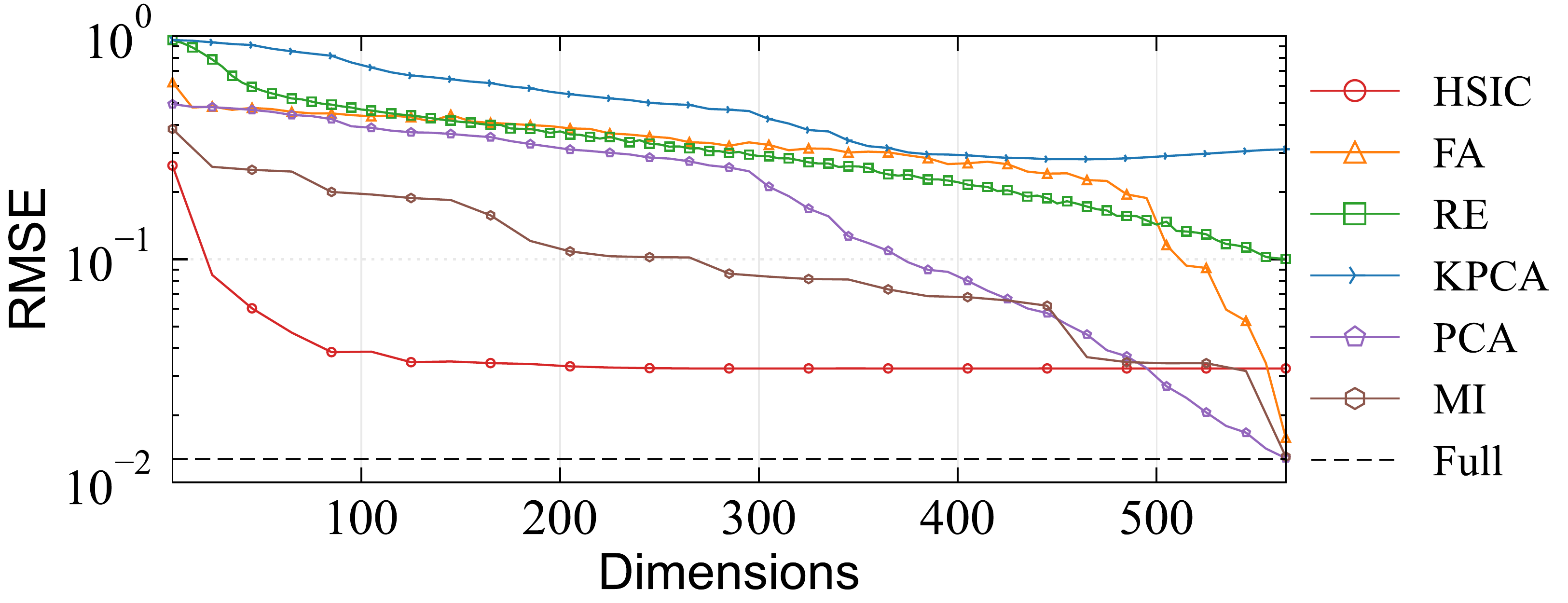}
    \caption{Feature reduction experiment}
    \Description{Feature reduction experiment}
    \label{dimension}
\end{figure} 

\section{Conclusion}
In this paper, we propose \ours, a shrinkage deep kernel learning to efficiently model high-dimensional input with an effective parallel batch updating scheme, to efficiently tackle the yield estimation of high-dimensional variational variables.
Compared to the SOTA methods, \ours shows nearly \guohao{10.3x} speed up with consistently accurate results. Limitation of \ours includes (1) solving entropy-based optimization relies on high-performance GPU and is time-consuming, and (2) the feature selection can ignore critical variables and lead to model bias for the yield estimation.

\balance %

\section{ACKNOWLEDGMENTS}
{This work is supported by the Fundamental Research Funds for the Central Universities. The experiment is supported by Primarius Technologies Co., Ltd.}

\bibliography{sample-base}
\end{document}